\documentclass[12pt]{article}

\usepackage{amsmath, amssymb, slashed}
\usepackage[dvips]{graphicx}
\usepackage{epsf}
\usepackage{color}
\usepackage{epstopdf}

\begin{document}
\def\a{\alpha}
\def\b{\beta}
\def\c{\varepsilon}
\def\d{\delta}
\def\e{\epsilon}
\def\f{\phi}
\def\g{\gamma}
\def\h{\theta}
\def\k{\kappa}
\def\l{\lambda}
\def\m{\mu}
\def\n{\nu}
\def\p{\psi}
\def\q{\partial}
\def\r{\rho}
\def\s{\sigma}
\def\t{\tau}
\def\u{\upsilon}
\def\v{\varphi}
\def\w{\omega}
\def\x{\xi}
\def\y{\eta}
\def\z{\zeta}
\def\D{{\mit \Delta}}
\def\G{\Gamma}
\def\H{\Theta}
\def\L{\Lambda}
\def\F{\Phi}
\def\P{\Psi}

\def\S{\Sigma}

\def\o{\over}
\def\beq{\begin{eqnarray}}
\def\eeq{\end{eqnarray}}
\newcommand{\gsim}{ \mathop{}_{\textstyle \sim}^{\textstyle >} }
\newcommand{\lsim}{ \mathop{}_{\textstyle \sim}^{\textstyle <} }
\newcommand{\vev}[1]{ \left\langle {#1} \right\rangle }
\newcommand{\bra}[1]{ \langle {#1} | }
\newcommand{\ket}[1]{ | {#1} \rangle }
\newcommand{\EV}{ {\rm eV} }
\newcommand{\KEV}{ {\rm keV} }
\newcommand{\MEV}{ {\rm MeV} }
\newcommand{\GEV}{ {\rm GeV} }
\newcommand{\TEV}{ {\rm TeV} }
\def\diag{\mathop{\rm diag}\nolimits}
\def\Spin{\mathop{\rm Spin}}
\def\SO{\mathop{\rm SO}}
\def\O{\mathop{\rm O}}
\def\SU{\mathop{\rm SU}}
\def\U{\mathop{\rm U}}
\def\Sp{\mathop{\rm Sp}}
\def\SL{\mathop{\rm SL}}
\def\tr{\mathop{\rm tr}}

\def\IJMP{Int.~J.~Mod.~Phys. }
\def\MPL{Mod.~Phys.~Lett. }
\def\NP{Nucl.~Phys. }
\def\PL{Phys.~Lett. }
\def\PR{Phys.~Rev. }
\def\PRL{Phys.~Rev.~Lett. }
\def\PTP{Prog.~Theor.~Phys. }
\def\ZP{Z.~Phys. }

\begin{titlepage}
\begin{center}

\hfill IPMU13-0212\\
\hfill ICRR-Report-665-2013-15\\
\hfill \today

\vspace{1.5cm}
{\large\bf 
Mixed (Cold+Warm) Dark Matter in the Bino-Wino co-annihilation scenario
}
\vskip 1.2cm
{ Masahiro Ibe}$^{(a,b)}$,
{ Ayuki Kamada}$^{(b)}$
and
{ Shigeki Matsumoto}$^{(b)}$\\

\vskip 0.4cm
{\it
$^{(a)}${\it ICRR, University of Tokyo, Kashiwa, 277-8583, Japan }\\
$^{(b)}${Kavli IPMU, University of Tokyo, Kashiwa, 277-8583, Japan} 
}

\vskip 1.5cm

\abstract{
We study phenomenological aspects of the bino-wino co-annihilation scenario in high-scale supersymmetry breaking models.
High-scale SUSY breaking scenarios are considered to be promising possibility after the discovery of the Higgs boson with a mass around $126$\,GeV.
In this paper, we discuss the bino lightest supersymmetric particle (LSP) accompanied by the at most around $30$\,GeV heavier wino.
With the suitable mass splitting between the bino and the wino, 
the bino LSP has the correct relic abundance of dark matter.
For the smaller mass splitting, the late-time decay of the gravitino 
can provide the correct abundance of the bino dark matter.
It is extremely challenging to find signals from the bino dark matter in direct and indirect detections.
By utilizing multi-jets plus missing transverse momentum events at the LHC, 
we can constraint the gluino mass and thus probe the bino-wino co-annihilation scenario indirectly.
The collider experiment, however, can not search the bino dark matter directly.
In this paper, we suggest the direct probe of the bino dark matter.
We show that the bino dark matter leaves imprints on the small-scale matter power spectrum 
when the bino dark matter is produced by the decay of the gravitino.
The non-thermal bino dark matter behaves as mixed (cold+warm) dark matter.
}

\end{center}
\end{titlepage}
\setcounter{footnote}{0}

\section{Introduction}
\label{sec: introduction}
It has remained a fundamental challenge to reveal the nature of dark matter, 
while the presence of dark matter has been confirmed in astrophysics and cosmology.
In particular, recent observations of the cosmic microwave background anisotropy\,\cite{wmap9, planck}
determine the relic density of dark matter almost at the several percent level,
\begin{eqnarray}
\Omega_{\rm dm}h^{2} \simeq 0.12 \,.
\end{eqnarray}
Our little but certain knowledge about the nature of dark matter, i.e. longevity and coldness, 
tells us that the standard model of particle physics does not provide a candidate of dark matter.
Therefore, the investigation of the nature and the origin of dark matter is important subject
in particle physics as well as astrophysics and cosmology.

The physics beyond the standard model, which is originally suggested to address the large hierarchy between the 
electro-weak scale and the scale of the grand unified theory (GUT) or the Planck scale, provide promising candidates of dark matter.
The lightest supersymmetric particle (LSP) in R-parity conserving supersymmetric (SUSY) theories is one of such candidates\,\cite{susydm}.
SUSY theories are also supported by the precise unification 
of the three gauge coupling constants of the standard model at the GUT scale.

SUSY models are reexamined after the discovery of the Higgs boson\,\cite{higgsatlas, higgscms} and null-detection of SUSY signals at the LHC.
Among them, high-scale SUSY breaking scenarios, which are characterised by the heavy gravitino with a mass of $m_{3/2} \simeq 10-1000$\,TeV,
attract considerable attention\,\cite{splitsusy, puregrav, spreadsusy, minimalSplit}.
In addition to explaining the relatively large Higgs mass, 
high-scale SUSY breaking scenarios evade the gravitino/Polony problems, the flavour changing neutral current problems, and the CP-problems.

In this paper, we consider the bino LSP with the at most around $30$\,GeV heavier wino as the next-to-LSP (NLSP) in high-scale SUSY breaking scenarios. 
Due to the heavy sfermions, the self-annihilation of the bino LSP is insufficient and its thermal relic over-closes the Universe if the slightly heavier wino is not accompanied.
On the other hand, the wino dark matter self-annihilates effectively, since the approximate custodial symmetry 
prohibits the large mass splitting between the neutral and the charged winos. 
The mass splitting between the neutral and the charged winos can be around $150-170$\,MeV\,\cite{winofeng, winogiudice, winomasssplitsato}.
The large annihilation cross section of the wino dark matter, however, is in tension with gamma-ray observations 
of the Galactic center in the Fermi-LAT and the H.E.S.S. telescope\,\cite{winogammacohen, winogammareece},
while there is still large ambiguity in the dark matter profile at the Galactic center.
These issues are solved if the mass splitting between the bino LSP and the wino NLSP is sufficiently small.
The small mass splitting keeps the wino and the bino abundance almost the same (except for the small Boltzmann factor) at the freeze-out. 
In this case, the efficient wino annihilation can reduce the resultant bino abundance to the observed dark matter density or below.
This mechanism is referred to the co-annihilation\,\cite{coannihilation}.

We also consider the late-time decay of the gravitino, which provides the non-thermal bino dark matter 
when the small mass splitting between the bino and the wino
does not leave the sufficient amount of the thermal bino dark matter.
The non-thermal binos are highly energetic at the time of the gravitino decay.
After that, they can loose their energy via interactions with the thermal background.
Since the bino does not interact with the standard model particles elastically, 
the energy-loss proceeds via a cycle of interactions with the winos.
The bino turns into the charged winos via the inelastic scattering by the thermal background.
This inelastic process triggers the energy-loss cycle of the bino.
The inelastic scattering rate is sensitive to the mass splitting between the bino and the wino, which
also determines the thermal relic abundance of the bino.
By direct integration of the Boltzmann equation with the energy-loss cycle, we show that a sizable fraction of the non-thermal bino 
remains relativistic after the energy-loss cycle becomes inefficient.
Therefore, the bino dark matter, which consists of thermal and non-thermal components, can be mixed (cold+warm) dark matter.
The imprints on the small-scale matter power spectrum may provide further insights on the origin of dark matter via the future $21{\rm cm}$ line survey\,\cite{21cmcosmology}. 

The organization of the paper is as follows.
In section\,\ref{sec:binowino}, we summarize the bino-wino co-annihiliation scenario
mainly assuming the pure gravity mediation\,\cite{puregrav}/minimal split SUSY model\,\cite{minimalSplit}, although
our discussion can be applied to generic bino-wino co-annihiliation scenarios.
In section\,\ref{sec:masssplitting}, we determines the mass splitting 
between the bino LSP and the wino NLSP for a given thermal relic abundance.
The mass splitting determines the inelastic scattering rate of the non-thermal bino.
In section\,\ref{sec:smallscale}, we discuss the imprints of the non-thermal bino
on the small-scale matter power spectrum, assuming that the non-thermal bino is 
produced by the decay of the gravitino.
Here, we clarify the energy-loss cycle of the non-thermal bino.
The final section is devoted to summary. 

\section{The bino-wino co-annihilation Scenario}
\label{sec:binowino}
First, we summarize the bino-wino co-annihilation scenario in high-scale SUSY breaking models. %
\footnote{
The bino-wino co-annihilation scenario, especially the thermal relic abundance of the bino, is studied 
in other contexts such as of the minimal supergravity model with non-universal gauging mass\,\cite{nonunivmixed, nonunivbwca}
and the gaugino condensation in hidden sector\,\cite{gauginocond}.
}
To be specific, we concentrate on the pure gravity mediation/minimal split SUSY model, 
although we do not need to change our discussion for generic bino-wino co-annihilation scenarios with the heavy sfermions and the heavy higgsinos.
Then, we discuss phenomenological implications of the model.

\subsection{Mass spectrum of the model}
In the pure gravity mediation/minimal split SUSY model, the sfermion mass $m_{\tilde f}$ and the higgsino mass $\mu$ originate from the generic tree-level interactions in the supergravity,
and thus they are expected to be of the order of the gravitino mass.
The gravitino mass is set $m_{3/2} \simeq 10-1000$\,TeV in order to explain the large quantum corrections to the Higgs mass.
On the other hand, the charge of the SUSY breaking field under some symmetry allows only one-loop suppressed contributions to the gaugino masses,
i.e. the anomaly mediated contributions\,\cite{anomalymedgiudice, anomalymedrandall}.
The heavy higgsinos also lead to large threshold effects to the gaugino masses, which are parametrized by $L$\,\cite{puregrav, winogiudice, anomalymedgiudice}.
The gaugino masses are numerically evaluated by
\begin{eqnarray}
\label{eq:gluinomass}
m_{\tilde g} &\simeq&
2.5\times (1 - 0.13 \, \delta_{3/2} - 0.04 \, \delta_{0})
\times 10^{-2} \, m_{3/2},
\\ 
\label{eq:winomass}
m_{\tilde w} &\simeq&
3.0\times(1 - 0.04 \, \delta_{3/2+L} + 0.02 \, \delta_{0})
\times 10^{-3} \, (m_{3/2} + L),
\\ 
\label{eq:binomass}
m_{\tilde b} &\simeq&
9.6\times(1 + 0.01 \, \delta_{0})
\times 10^{-3} \, (m_{3/2} + L/11),
\end{eqnarray}
where the subscripts $\tilde g$, $\tilde w$ and $\tilde b$ denote gluino, wino and bino, respectively.
Here, $\delta_{0} = \log[m_{\tilde f}/100 \,{\rm TeV}]$, 
$\delta_{3/2} = \log[m_{3/2}/100 \, {\rm TeV}]$, 
and $\delta_{3/2+L} = \log[(m_{3/2} + L) /100 \,{\rm TeV}]$. 
The terms proportional to $m_{3/2}$ and $L$ in the above formulas represent the anomaly mediation contributions 
and the higgsino threshold effects, respectively.

In the following discussion, we set the sfermion mass and the higgsino mass equal to the gravitino mass $m_{\tilde f}=\mu=m_{3/2}$.
The above formulas of the gaugino masses imply that bino is LSP for $L/m_{3/2} > 3$.
We focus on the parameter region of $L/m_{3/2} \simeq 3.5-4.0$, 
where the mass splitting between the bino and the wino, ${\mit \Delta} m_{\tilde b} \equiv m_{\tilde w} - m_{\tilde b}$, 
is about $10\,\%$ of the bino mass and the bino-wino co-annihilation is efficient.
In the next section, we identify the suitable mass splitting for a given thermal relic density.

\subsection{Phenomenological aspects of the model}
\label{subsec:phenomenology}
Here, we describe phenomenological implications of the heavy sfermions and the heavy higgsinos.
First, the large $\mu$-term suppresses the bino-wino mixing to ${\mathcal O}(m_{Z}^{2} / (\mu |m_{\tilde w} - m_{\tilde b}|))$.
Even with the small mass splitting between the bino and the wino,
the bino compose at least $0.99$ of LSP in the parameter region of interest. 
The tiny mixing and the heavy sfermions result in the extremely weakly interacting bino.
Therefore, in the present model, the bino dark matter does not leave detectable signals 
in current and near-future direct and indirect detections of dark matter.

The only constraint is put by null-detection of the gluino signals at the LHC.
By using multi-jets plus missing transverse momentum events, the ATLAS experiment reports
that the gluino mass should be $m_{\tilde g} \gtrsim 1.2$\,TeV for the LSP mass below $500$\,GeV\,\cite{atlasgluino}.
In the near future, the $14$\,TeV LHC with $300$\,fb$^{-1}$ has a potential to discover the gluino 
below $m_{\tilde g} \simeq 2.3$\,TeV\,\cite{nearfuturegluino}.
Furthermore, a $33$\,TeV future proton collider with $3000$\,fb$^{-1}$ can reach the gluino mass of $m_{\tilde g} \simeq 3.6$\,TeV\,\cite{futuregluino}.
Here, we should note that in these collider experiments, we can probe only the gluino production and its subsequent cascade decay, but not the bino LSP itself.
This motivates us to study the cosmological imprints of the bino dark matter as the direct probe for the present scenario in this paper.

The large $\mu$-term also ensures the small mass splitting between the neutral wino 
${\tilde w}^{0}$ and the charged winos ${\tilde w}^{\pm}$, 
${\mit \Delta} m_{\tilde w} \equiv m_{{\tilde w}^{\pm}} - m_{{\tilde w}^{0}}$. 
This is because the large $\mu$-term suppresses the effect of the approximate custodial symmetry on the mass splitting ${\mit \Delta} m_{\tilde w}$.
The tree-level contribution to the mass splitting is ${\mit \Delta} m^{\rm tree}_{\tilde w} \lesssim 20$\,MeV
in the parameter region of interest.
On the other hand, the one-loop contribution\,\cite{winofeng, winogiudice} is given by
\begin{eqnarray}
\label{eq:deltaM}
{\mit \Delta} m^{\rm loop}_{\tilde w}  = m_{\tilde w^\pm}- m_{\tilde w^0} =
\frac{g_2^2}{16\pi^2} m_{\tilde w}
\left[ f(r_W) - \cos^2 \theta_W f(r_Z) - \sin^2 \theta_W f(0) \right],
\end{eqnarray}
where $\theta_{W}$ is the Weinberg angle, $f(r)= \int^1_0 dx (2 + 2 x^2) \ln[x^2 + (1 - x)r^2]$, and $r_{W,Z}=m_{W,Z}/m_{\tilde w}$ with the $W,Z$-boson mass $m_{W,Z}$.
The one loop contribution is ${\mit \Delta} m^{\rm loop}_{\tilde w} \simeq 150-170$\,MeV, %
\footnote{
Recently the mass splitting is carefully analysed up to the two-loop contribution
in the similar context of the high scale SUSY breaking model\,\cite{winomasssplitsato}.
The two loop contribution can shift the mass splitting at most $10$\,MeV.
We ignore this small effect, which does not change our discussion.
}
and thus dominates the mass splitting ${\mit \Delta} m_{\tilde w}$.
In the other scenarios with $|m_{\tilde b}-m_{\tilde w}| \sim m_{\tilde b}\,(m_{\tilde w})$, 
the tree-level contribution is negligible compared with the one-loop contribution,
${\mit \Delta} m^{\rm tree}_{\tilde w} < {\mathcal O}(100)$\,keV.
However, in the present bino-wino co-annihilation scenario ${\mit \Delta} m_{\tilde b}/m_{\tilde b} \ll 1$,
the tree-level contribution is subdominant but can not be ignored, ${\mit \Delta} m^{\rm tree}_{\tilde w}/{\mit \Delta} m^{\rm loop}_{\tilde w} \lesssim 0.1$.


\section{Mass splitting between bino and wino}
\label{sec:masssplitting}
The interactions of the bino LSP is extremely weak due to the heavy sfermions and the heavy higgisnos.
The inefficient annihilation of the bino LSP generically leads to the over-closure of the Universe.
However, when the bino is accompanied by the slightly heavier wino, the presence of the wino at the freeze-out of bino reduces
the relic abundance of the bino.
For the heavier bino with $m_{\tilde b} \gtrsim 1$\,TeV, the bino-wino co-annihilation is also boosted up 
by the non-perturbative effects.
The countless exchanges of gauge bosons ($\gamma, W, Z$) between the winos distort the initial state wave function 
from the direct product of the plane waves.
This is known as the Sommerfeld enhancement\,\cite{sommerfeldgeneral}.
In this section, we calculate the bino relic abundance taking into account the co-annihilation and the Sommerfeld enhancement.

\subsection{Thermally averaged effective cross section}
The thermal relic of the non-relativistic stable particle can be evaluated by solving the following Boltzmann equation\,\cite{thermalrelic},
\begin{eqnarray}
\label{eq:boltzmannchemical}
\frac{dY}{dx}=-\frac{\langle \sigma v \rangle}{H x} \left( 1-\frac{x}{3 g_{*s}} \frac{dg_{*s}}{dx} \right)
s \, (Y^{2} - Y^{2}_{\rm eq}) \,,
\end{eqnarray}
with the time coordinate $x$, which is the inverse of the cosmic temperature $T$ normalized by the particle mass $m$, $x \equiv m/T$.
The yield of the particle $Y$ is defined by the ratio of the particle number density $n$ to the entropy density $s$, $Y=n/s$.
The equilibrium yield $Y_{\rm eq}$ is given by,
\begin{eqnarray}
Y_{\rm eq} = \frac{g}{(2 \pi)^{3/2}} \frac{m^{3}}{s \, x^{3/2}} e^{-x} \,,
\end{eqnarray}
with the degrees of freedom of the particle $g$.
The entropy density and the cosmic expansion rate $H$ is given by,
\begin{eqnarray}
s = g_{*s} \frac{2 \pi^{2}}{45} \frac{m^{3}}{x^{3}} \,, \quad
H = \left( \frac{g_{*}}{10} \right)^{1/2} \frac{\pi}{3 M_{\rm pl}} \frac{m^{2}}{x^{2}} \,,
\end{eqnarray}
with the reduced Planck mass $M_{\rm pl} \simeq 2.43 \times 10^{18}$\,GeV.
The effective degrees of freedom for the energy density $g_{*}$ and for the entropy density $g_{*s}$ is calculated as the function of $x$ 
according to Ref.\,\cite{effectivedegrees}.
The thermally averaged annihilation cross section $\langle \sigma v \rangle$ is
related to the annihilation cross section $(\sigma v)$ by,
\begin{eqnarray}
\langle \sigma v \rangle = \left( \frac{m}{4 \pi T} \right)^{3/2} \int 4 \pi v^{2} dv \, (\sigma v) \, \exp \left( -\frac{m v^{2}}{4T} \right) \,,
\end{eqnarray}
with the relative velocity of the initial particles, $v$.

The co-annihilation introduces two changes in the above formulas\,\cite{winorelic},
\begin{eqnarray}
g &\to& g_{\rm eff} =  \sum_{i}  g_{i}(1+{\mit \Delta}_{i})^{3/2} e^{-x {\mit \Delta}_{i}} \,, \\
\label{eq:coannihilation}
\langle \sigma v \rangle &\to& \langle \sigma_{\rm eff} v \rangle = \sum_{i, j} \langle \sigma_{i j} v \rangle \frac{g^{2}}{g_{\rm eff}^2}
(1+{\mit \Delta}_{i})^{3/2} (1+{\mit \Delta}_{j})^{3/2} e^{- x({\mit \Delta}_{i}+{\mit \Delta}_{j})} \,.
\end{eqnarray}
The index $i, j$ runs over the all particles that contribute to the co-annihilation including the stable particle itself.
The dimensionless mass splitting between the stable particle and the particle $i$ is defined by ${\mit \Delta}_{i} = m_{i}/m - 1$.
Since the thermal relic of the stable particle freeze-out at $x \simeq 20$ and 
the thermally averaged effective annihilation cross section $\langle \sigma_{\rm eff} v \rangle$ 
depends exponentially on the dimensionless mass splitting (Eq.\,(\ref{eq:coannihilation})) , 
the co-annihilation is significant
only with at most $\sim 10 \, \%$ heavier particles than the stable particle.

The above formulas of the co-annihialation are valid only if the chemical equilibrium between the stable particle and the accompanying particles 
is kept around the freeze-out.
In order to check this point, we should specify the inelastic interactions in charge of keeping the chemical equilibrium.
In the case of our present bino-wino co-annihilation scenario, the dominant processes are
${\tilde b} + e\,(\nu_{e}) \leftrightarrow {\tilde w}^{\pm} + \nu_{e}\,(e)$, ${\tilde w}^{0} + e\,(\nu_{e}) \leftrightarrow {\tilde w}^{\pm} + \nu_{e}\,(e)$,
${\tilde w}^{\pm} \to {\tilde b} + f_{1} + {\bar f}_{2}$, ${\tilde w}^{\pm} \to {\tilde w}^{0} + \pi$ and ${\tilde w}^{0} \to {\tilde b} + f + {\bar f}$.
The $2$-body decays of ${\tilde w}^{\pm} \to {\tilde b} + \pi$ and ${\tilde w}^{0} \to {\tilde b} + \pi$ are sub-dominant
compared with the $3$-body decays for $\Delta m > 1$\,GeV\,\cite{2decay}.
The interaction rates are given by,
\begin{eqnarray}
&& \Gamma_{{\tilde b} + e\,(\nu_{e}) \to {\tilde w}^{\pm} + \nu_{e}\,(e)} = 4 N_{1 2}^{2}
\langle \Gamma_{\rm fi} (p_{\tilde b}, T; m_{\tilde b}, {\mit \Delta} m_{\tilde b}+{\mit \Delta} m_{\tilde w}) \rangle_{\tilde b} \,, \\
\label{eq:winononrelainelastic}
&& \Gamma_{{\tilde w}^{0} + e\,(\nu_{e}) \to {\tilde w}^{\pm} + \nu_{e}\,(e)} = 4
\langle \Gamma_{\rm fi} (p_{\tilde w}, T; m_{\tilde w}, {\mit \Delta} m_{\tilde w}) \rangle_{{\tilde w^{0}}} \,, \\
&& \Gamma_{{\tilde w}^{\pm} + \nu_{e}\,(e) \to {\tilde b} + e\,(\nu_{e})} = 4 N_{1 2}^{2}
\langle  \Gamma_{\rm bi} (T; m_{\tilde w}) \rangle_{{\tilde w^{\pm}}} \,, \\
&& \Gamma_{{\tilde w}^{\pm} + \nu_{e}\,(e) \to {\tilde w}^{0} + e\,(\nu_{e})} = 4
\langle  \Gamma_{\rm bi} (T; m_{\tilde w}) \rangle_{{\tilde w^{\pm}}} \,, \\
&& \Gamma_{{\tilde w}^{\pm} \to {\tilde b} + f_{1} + {\bar f}_{2}} = N_{1 2}^{2} {\bigg [} \sum_{l} \Gamma_{c} (m_{{\tilde w}^{\pm}}, m_{\tilde b}, m_{l}, 0) \notag \\
&& \qquad \qquad \qquad \quad
+  3\sum_{U, D} \left| V^{\rm CKM}_{U D} \right|^{2} \Gamma_{c} (m_{{\tilde w}^{\pm}}, m_{\tilde b}, m_{q_{1}}, m_{q_{2}}) {\bigg ]} \,, \\
&& \label{eq:2-body}
 \Gamma_{{\tilde w}^{\pm} \to {\tilde w}^{0} + \pi} = \frac{2}{\pi} \left| V^{\rm CKM}_{u d} \right|^{2}  G_{F}^{2} f_{\pi}^{2} 
{\mit \Delta}m_{\tilde w}^{3} \left( 1 - \frac{m_{\pi}}{{\mit \Delta}m_{\tilde w}} \right)^{1/2}   \,,\\
&& \Gamma_{{\tilde w}^{0} \to {\tilde b} + f + {\bar f}} = {\bigg [} \frac{1}{2} \left ( N_{1 3} - N_{2 3} \tan\theta_{W} \right) \cos\beta
- \frac{1}{2} \left ( N_{1 4} - N_{2 4} \tan\theta_{W} \right) \sin\beta {\bigg ]}^{2} \notag \\
&& \qquad \qquad \qquad
\times {\bigg [} \sum_{l}  \Gamma_{n} (m_{{\tilde w}^{0}}, m_{\tilde b}, m_{l}) + 3 \sum_{q}  \Gamma_{n} (m_{{\tilde w}^{0}}, m_{\tilde b}, m_{q}) {\bigg ]} \notag \\
&& \qquad \qquad \qquad
+ {\bigg [} \left( \frac{1}{2} \right)^{2} \sum_{l}  \Gamma_{\tilde f} (m_{{\tilde w}^{0}}, m_{\tilde b}, m_{l}) 
\notag \\
&& \qquad \qquad \qquad \qquad
+ 3 \left( \frac{1}{6} \right)^{2} \sum_{q}  \Gamma_{\tilde f} (m_{{\tilde w}^{0}}, m_{\tilde b}, m_{q}) {\bigg ]}
\,,
\end{eqnarray}
with the Fermi constant $G_{F}$, the pion mass $m_{\pi}$, and the pion decay constant $f_{\pi}$.
The mixing matrix of the neutralinos $N_{i j}$ is defined in the same way as in Ref.\,\cite{susyprimer}.
The indices $(i, j)$ on $N_{i j}$ correspond to (mass, gauge) eigenstates.
The subscripts $q$s and $l$s represent the quarks and the leptons, respectively.
The Cabbibo-Kobayashi-Maskawa matrix is denoted by $V^{\rm CKM}_{U D}$.
The indices $(U, D)$ on $V^{\rm CKM}_{U D}$ correspond to (up, down)-type quarks, 
while the indices $(u, d)$ on $V^{\rm CKM}_{u d}$ in Eq.\,(\ref{eq:2-body}) denote the (up, down) quarks concretely.
The ratio of the vacuum expectation values of the (up, down)-type Higgs, $(v_{u}, v_{d})$ is written as $\tan \beta = v_{u} / v_{d}$.
It should be noted that $p_{{\tilde b}, {\tilde w}}$ is the physical momentum of the bino (wino) at the cosmic temperature $T$.
Here, we define
\begin{eqnarray}
&& \Gamma_{\rm fi} (p, T; m, {\mit \Delta} m) = \frac{4}{3 \pi^{3}} G_{F}^{2} T^{5} 
{\bigg [} 72 + 36 \frac{{\mit \Delta} m}{T} + 6 \left( \frac{{\mit \Delta} m}{T} \right)^{2} \notag \\
&& \qquad \qquad \qquad \qquad \quad 
- 6 \frac{p}{E} \left( \frac{{\mit \Delta} m}{T} \right)^{3} 
+ 4 \left( \frac{p}{E} \right)^{2} \left( \frac{{\mit \Delta} m}{T} \right)^{4} {\bigg ]} 
\exp \left( - \frac{{\mit \Delta} m}{T} \right) \,, \\
&& \Gamma_{\rm bi} (p, T; m) =  \frac{96}{\pi^{3}} G_{F}^{2} T^{5} 
\frac{E^{2}+p^{2}}{m^{2}} \,, \\
&& \Gamma_{c} (M_{1}, M_{2}, m_{1}, m_{2}) = \frac{1}{3 \pi^{3}} G_{F}^{2}  \int_{M_{2}}^{\frac{M_{1}^{2} + M_{2}^{2} - (m_{1} + m_{2})^{2}}{2 M_{1}}} 
dE_{2} \, p_{2} \left(1 - \frac{q^{2}}{m_{W}^{2}} \right)^{-2} \notag \\
&& \qquad \quad
\times {\bigg [} C_{1} \left( 2 (M_{1} E_{2} - M_{2}^{2}) (M_{1} - E_{2}) - M_{2} q^2 \right)
- C_{2} \left( 2 M_{2} - E_{2} \right) q^{2} {\bigg ]} \\
&& \qquad \qquad \qquad \quad
\stackrel{M_{1} \to M_{2},\,m_{1}=m_{2}=0}{\to} \frac{2}{15 \pi^{3}} G_{F}^{2} (M_{1} - M_{2})^{5} \,, \\
&& \Gamma_{n} (M_{1}, M_{2}, m) = \frac{1}{64 \pi^{3}} \frac{g_{2}^{4}}{m_{h}^{4}} \left( \frac{m}{m_{W}} \right)^{2} \int_{M_{2}}^{\frac{M_{1}^{2} + M_{2}^{2} - 4m^{2}}{2 M_{1}}}  
dE_{2} \, p_{2} \left(1 - \frac{q^{2}}{m_{h}^{2}} \right)^{-2} \notag \\
&& \qquad \qquad \qquad \qquad
\times  \left(1 - \frac{4 m^{2}}{q^{2}} \right)^{3/2} q^{2} \left( M_{2} + E_{2} \right) \\
&& \qquad \qquad \qquad
\stackrel{M_{1} \to M_{2},\,m=0}{\to} \frac{1}{240 \pi^{3}} \frac{g_{2}^{4}}{m_{h}^{4}} \left( \frac{m}{m_{W}} \right)^{2} (M_{1} - M_{2})^{5} \,, \\
&& \Gamma_{\tilde f} (M_{1}, M_{2}, m, c_{{\tilde b}{\tilde w}}) = \frac{1}{320 \pi^{3}} \frac{g_{1}^{2} g_{2}^{2}}{m_{\tilde f}^{4}} \int_{M_{2}}^{\frac{M_{1}^{2} + M_{2}^{2} - 4m^{2}}{2 M_{1}}} 
dE_{2} \, p_{2}  \notag \\
&& \qquad \qquad \qquad \qquad \qquad
\times {\bigg [} C_{1} \left( 2 (M_{1} E_{2} - M_{2}^{2}) (M_{1} - E_{2}) + c_{{\tilde b}{\tilde w}} M_{2} q^2 \right) \notag \\
&& \qquad \qquad \qquad \qquad \qquad \qquad
+ C_{2} \left( 2 c_{{\tilde b}{\tilde w}} M_{2} + E_{2} \right) q^{2} {\bigg ]} \notag \\
&& \qquad \qquad \qquad \qquad
\stackrel{M_{1} \to M_{2},\,m=0}{\to} \frac{(2 + c_{{\tilde b}{\tilde w}})}{800 \pi^{3}} \frac{g_{1}^{2} g_{2}^{2}}{m_{\tilde f}^{4}} (M_{1} - M_{2})^{5} \,, \\
&& C_{1} = \left( q^{2} - (M_{1} + M_{2})^{2} \right)^{1/2} \left( q^{2} - (M_{1} - M_{2})^{2} \right)^{1/2} \notag \\
&& \qquad \quad
\times \left( q^{4} + (M_{1}^{2} + M_{2}^{2}) q^{2} - 2 (M_{1}^{2} - M_{2}^{2})^{2}  \right) / q^{6} \,, \\
&& C_{2} = \left( q^{2} - (M_{1} + M_{2})^{2} \right)^{3/2} \left( q^{2} - (M_{1} - M_{2})^{2} \right)^{3/2} / q^{6} \,, \\
&& q^{2} = M_{1}^{2} + M_{2}^{2} - 2 M_{1} E_{2} \,,
\end{eqnarray}
with the mass of the standard model Higgs boson $m_{h}$.
We normalize $g_{1}$ and $g_{2}$ such that $g_{1} = \sqrt{5/3} \, g'$ and $g_{2} = g$ 
with the conventional electro-weak gauge couplings $g$ and $g'$ ($e = g \sin \theta_{W} = g' \sin \theta_{W}$ with the positron charge $e$).
The quantity in angle brackets $\langle Q \rangle_{\tilde b,\,(\tilde w)}$ denotes the average of quantity $Q$ over the bino (wino) thermal distribution.
The relative phase between the bino and the wino mass parameter ${c_{{\tilde b}{\tilde w}}}$ is unity in the present model, ${c_{{\tilde b}{\tilde w}}}=1$.

\begin{figure}[t]
 \begin{center}
 \includegraphics[width=0.6\linewidth]{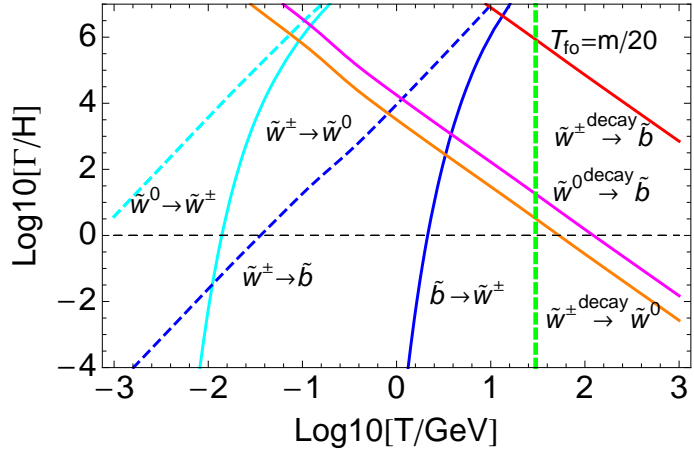}
 \caption{\sl \small The reaction rates of the inelastic processes in charge of keeping the chemical equilibrium between the bino $\tilde b$ and the winos $\tilde w$s.
In the explanation of each line, we omit the standard model particles and explicitly specify the decay processes. Here, we set $m_{\tilde b} = 600$\,GeV and ${\mit \Delta} m_{\tilde b} = 31.5$\,GeV.}
 \label{fig:chemical1}
 \end{center}
\end{figure}

In Fig.\,\ref{fig:chemical1}, we show the reaction rates for $m_{\tilde b} = 600$\,GeV and ${\mit \Delta} m_{\tilde b} = 31.5$\,GeV.
The inelastic scatterings are efficient to keep the chemical equilibrium between $\tilde b$, ${\tilde w}^{0}$, and ${\tilde w}^{\pm}$ until the freeze-out of the bino LSP.
It should also be noted that the 3-body decay of ${\tilde w}^{\pm} \to {\tilde b} + f_{1} + f_{2}$, 
not the 2-body decay of ${\tilde w}^{\pm} \to {\tilde w}^{0} + \pi$, dominates the decay of the charged winos.

The Sommerfeld enhancement is incorporated for the annihilations of the winos by calculating the annihilation cross section as
\begin{eqnarray} 
\sigma_{i j} v = c_{i j} \sum_{(k, l), (m,n)} d_{i j, k l} d^{*}_{m n, i j} \Gamma_{k l, m n} \,,
\end{eqnarray}
where $c_{i i} = 2$ and $c_{i j} = 1$ ($i \neq j$), and $\Gamma_{i j, k l}$ is the absorptive term between two body states $\Phi_{i j}$ and $\Phi_{k l}$
\,\cite{sommerfeldgeneral}.
The enhancement factor $d_{i j, k l}$ is given by the distortion of the wave function with respect to the relative position $r$ of the annihilation particles
at the infinity,
\begin{eqnarray}
g^{>}_{i j, k l} (r) \stackrel{r \to \infty}{\to} d_{i j, k l} \exp(i \sqrt{mE} r) \,,
\end{eqnarray}
where $g^{>}_{i j, k l} (r)$ is the solution of the Schr{\" o}dinger equation of
\begin{eqnarray}
\label{eq:schrodinger}
-\frac{1}{m} \frac{d^{2}}{dr^{2}} g^{>}_{i j, k l} (r) + \sum_{(m, n)} V_{i j, m n} \, g^{>}_{m n, k l} (r) = E g^{>}_{i j, k l} (r) \,,
\end{eqnarray}
with the potential $V_{i j, k l} (r)$ and the boundary condition of 
\begin{eqnarray}
g^{>}_{i j, k l} (0) = 0 \,, \quad g^{>}_{i j, k l} (r) \stackrel{r \to \infty}{\propto} \exp(i \sqrt{mE} r) \,.
\end{eqnarray}
The absorptive term and the potential for each annihilation channel of the winos are summarized in Table\,\ref{table:winoannihilation}.

\begin{table}[tb]
 \caption{\sl \small
 The summary of absorptive term $\Gamma_{i j, k l}$ and the potential $V_{i j, k l} (r)$ for each annihilation channel of the winos.
 Here, we take an abbreviated notation of $c_{W} = \cos \theta_{W}$, $\alpha=e^{2}/4\pi$, and $\alpha_{2}=g_{2}^{2}/4\pi$.}
 \label{table:winoannihilation}
  \begin{center}
  \begin{tabular}{|c||c|c|} \hline
  channel ($i$, $j$) $\leftrightarrow$ ($k$, $l$) & $\Gamma_{i j,kl}$ &  $V_{i j,kl} (r)$  \\ \hline \hline
  (${\tilde \chi}^{+ (-)}$, ${\tilde \chi}^{+ (-)}$) (S=0) 
  & ${\displaystyle \frac{\pi \alpha_{2}^{2}}{2 m^{2}}}$ 
  & $\alpha {\displaystyle \frac{1}{r}} + \alpha_{2} c_{W}^{2} {\displaystyle \frac{e^{-m_{Z} r}}{r}}$ \\ \hline
  (${\tilde \chi}^{0}$, ${\tilde \chi}^{\pm}$) (S=0, 1)  
  & ${\displaystyle \frac{\pi \alpha_{2}^{2}}{2 m^{2}}}, {\displaystyle \frac{25 \pi \alpha_{2}^{2}}{24 m^{2}}}$ 
  & $-\alpha_{2} {\displaystyle \frac{e^{-m_{W} r}}{r}}$ \\ \hline
  (${\tilde \chi}^{+}$, ${\tilde \chi}^{-}$) (S=1)  
  & ${\displaystyle \frac{25 \pi \alpha_{2}^{2}}{24 m^{2}}}$
  & $-\alpha {\displaystyle \frac{1}{r}} - \alpha_{2} c_{W}^{2} {\displaystyle \frac{e^{-m_{Z} r}}{r}}$ \\ \hline
  (${\tilde \chi}^{+}$, ${\tilde \chi}^{-}$) $\leftrightarrow$ (${\tilde \chi}^{0}$, ${\tilde \chi}^{0}$) (S=0)
  & ${\displaystyle \frac{\pi \alpha_{2}^{2}}{2 m^{2}}}
        \begin{pmatrix}
        3 & \sqrt{2} \\
        \sqrt{2} & 3
        \end{pmatrix}$
  & $\begin{pmatrix}
        2 {\mit \Delta}m_{\tilde w} -\alpha {\displaystyle \frac{1}{r}} - \alpha_{2} c_{W}^{2} {\displaystyle \frac{e^{-m_{Z} r}}{r}} 
        & -\sqrt{2} \alpha_{2} {\displaystyle \frac{e^{-m_{W} r}}{r}} \\
        -\sqrt{2} \alpha_{2} {\displaystyle \frac{e^{-m_{W} r}}{r}} 
        & 0
        \end{pmatrix}$ \\ \hline
  \end{tabular}
 \end{center}
\end{table}

\subsection{The mass splitting between the bino and the wino ${\mit \Delta}m_{\tilde b}$}
\begin{figure}[t]
 \begin{center}
 \includegraphics[width=0.6\linewidth]{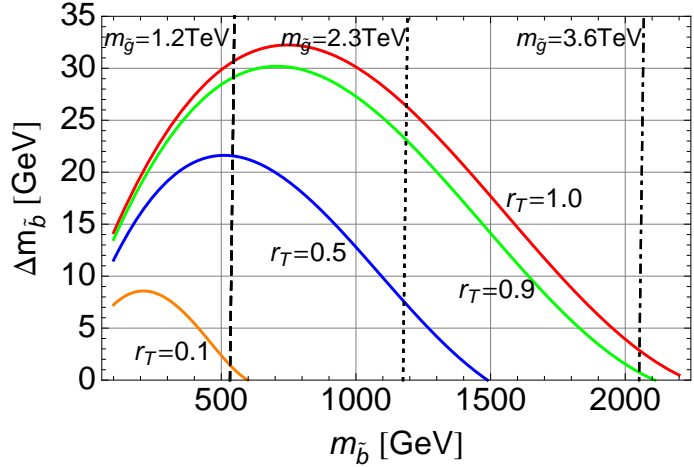}
 \caption{\sl \small 
 The mass splitting ${\mit \Delta}m_{\tilde b}$ as a function of the bino mass.
 Each line represents the thermal abundance of $r_{\rm T} = 1.0, \, 0.9, \, 0.5, \, 0.1$.
 The vertical lines show the constraints on the bino mass from the gluino search in the current LHC (dashed), the near-future LHC (dotted), and the future high-energy proton collider (dot-dashed).
 }
 \label{fig:rt}
 \end{center}
\end{figure}

We numerically solve the Boltzmann equation (Eq.\,(\ref{eq:boltzmannchemical})) with the co-annihilation and the Sommerfeld enhancement.
We identify the suitable mass splitting between the bino and the wino ${\mit \Delta}m_{\tilde b}$ for a given bino thermal relic, 
$r_{\rm T} \equiv \Omega_{\tilde b}^{\rm T} / \Omega_{\rm dm} = 1.0, \, 0.9, \, 0.5, \, 0.1$, up to $5\,\%$.
The results are shown in Fig.\,\ref{fig:rt}.

Here, we comment on the resonance in the Sommerfeld enhancement that is induced by the existence of the zero-energy binding states.
As shown in Ref.\,\cite{winorelic}, the wino relic abundance is resonantly reduced around some critical mass of $m_{\tilde w} > 2$\,TeV.
The resonance may change the monotonic decrease of ${\mit \Delta}m_{\tilde b}$ in large $m_{\tilde w}$ (see Fig.\,\ref{fig:rt}). 
The first resonance appears around $m_{\tilde b} \simeq 2.4$\,TeV in our numerical calculation in the case of ${\mit \Delta}m_{\tilde b} = 0$.
However, we find that with a tiny mass splitting of at most ${\mit \Delta}m_{\tilde b} \simeq 10$\,MeV, the bino thermal relic can explain the observed dark matter density 
even around $m_{\tilde b} \simeq 2.4$\,TeV.
This is because the resonance is highly sensitive to the freeze-out temperature, 
which depends on the mass splitting between the bino and the wino ${\mit \Delta}m_{\tilde b}$ (see Eq.\,(\ref{eq:coannihilation})).
The tiny mass splitting leads to an order of one mixing between the bino and the wino, i.e. the bino-wino dark matter rather than the bino dark matter.
Since the phenomenology of the bino-wino dark matter is not the subject of this paper, 
in the following we restrict ourselves within $m_{\tilde b} \lesssim 2.4$\,TeV. 

The mass splitting ${\mit \Delta}m_{\tilde b}$ for $r_{\rm T} = 1.0$, plotted in Fig.\,\ref{fig:rt}, is the upper limit.
If the mass splitting is larger, the bino thermal relic over-closes the Universe.
On the other hand, for $r_{\rm T} < 1$, the thermal relic can not explain the observed mass density of dark matter.
In that case, we need another production mechanism of the bino after the thermal freeze-out.
We assume that the late-time decay of the gravitino is in charge.
In the present scenario, the gravitino decays into the gauginos before the Big Bang Nucleosynthesis (BBN), 
\begin{eqnarray}
\label{eq:decaytemp}
T_{\rm d} = \left( \frac{10}{\pi^{2} g_{*}} M_{\rm pl}^{2} \Gamma_{3/2}^{2} \right)^{1/4} \simeq 0.12 \,{\rm MeV} \left( \frac{m_{3/2}}{10\,{\rm TeV}} \right)^{3/2} \,.
\end{eqnarray}
Here, the decay rate of the gravitino is given by,
\begin{eqnarray}
\Gamma_{3/2} \simeq \frac{8+3+1}{32 \pi} \frac{m_{3/2}^{3}}{M_{\rm pl}^{2}}\,.
\end{eqnarray}
The yield of the gravitino depends on the cosmic reheating temperature 
$T_{\rm R}$.
The resultant non-thermal bino relic is given by\,\cite{gravitinoabundance}, 
\begin{eqnarray}
\Omega_{\tilde b}^{\rm NT}h^{2} \simeq 2.7 \times 10^{-2} \left( \frac{m_{\tilde b}}{100\,{\rm GeV}} \right) \left( \frac{T_{\rm R}}{10^{10}\,{\rm GeV}} \right) \,.
\end{eqnarray}

The non-thermal bino produced by the late time decay of the gravitino is relativistic.
The dark matter with sizable peculiar velocity is called hot or warm dark matter, depending on the comoving velocity (roughly speaking, warm dark matter has comoving velocity of $v/c \sim 10^{-(7-8)}$).
The peculiar velocity of warm dark matter suppresses the growth of primordial perturbations and leaves the cut-off in the matter power spectrum
around the galactic or the sub-galactic scales\,\cite{warmdarkmatter}.
The cut-off in the matter power spectrum is not only characteristic feature of the nature of dark matter, 
but also possible solution to the so-called ``small-scale crisis"\,\cite{cuspcore, missingsatellite, reviewkravtsov, toobigtofail}.
As we will see in the next section, in the present scenario, the bino dark matter is mixed (warm+cold) dark matter rather than pure warm dark matter in the favored parameter range. 
However, tomography of the matter density fluctuations in the future $21$\,cm line observations is expected to give us a chance to find
even weaker imprints on the matter power spectrum\,\cite{21cmsmallscale}.
As mentioned in subsection\,\ref{subsec:phenomenology}, it is highly challenging to find the bino dark matter in direct and indirect detections.
Therefore, the imprints on the matter power spectrum can give valuable evidence of the bino-wino co-annihilation scenario
in high-scale SUSY models.


\section{Imprints on the small-scale matter power spectrum}
\label{sec:smallscale}
In this section, we study the imprints of the non-thermal bino on the small-scale matter power spectrum.
The comoving velocity of the non-thermal bino at the gravitino decay can be estimated as,
\begin{eqnarray}
\label{eq:comovingvb}
v_{\tilde b} / c \simeq 6.8 \times 10^{-8} \left( \frac{m_{\tilde b}}{100\,{\rm GeV}} \right)^{-1} \left( \frac{m_{3/2}}{10\,{\rm TeV}} \right)^{-1/2} \,.
\end{eqnarray}
The non-thermal bino is sufficiently energetic to be warm dark matter when they are produced.
However, after that, they may lose their energy through the interactions with the thermal background.
Our goal is to obtain the momentum spectrum of the bino dark matter after the energy-loss processes become inefficient.
To this end, first, we clarify the dominant energy-loss process for non-thermal bino.
Then, we derive and solve the Boltzmann equation of the momentum spectrum of the ``warm'' bino dark matter. 
Finally, we introduce two quantities that characterize the imprints of mixed dark matter on the matter power spectrum,
and calculate them from the obtained momentum spectrum.

\subsection{Energy-loss process}
As mentioned above repeatedly, the bino LSP in the present scenario does not elastically interact 
with the standard model particles, i.e. thermal background.
On the other hand, the winos can be a messenger between the bino and the standard model particles due to the small mass splitting 
${\mit \Delta} m_{\tilde b}$.
The energy-loss of the non-thermally produced wino is investigated for the wino LSP in previous works\,\cite{winoenergylossarcadi, winoenergylossibe}.
Here we summarize their results, and discuss points of modification when we apply the previous results to the present scenario.

\begin{figure}[tb]
 \begin{minipage}{.49\linewidth}
 \begin{center}
 \includegraphics[width=\linewidth]{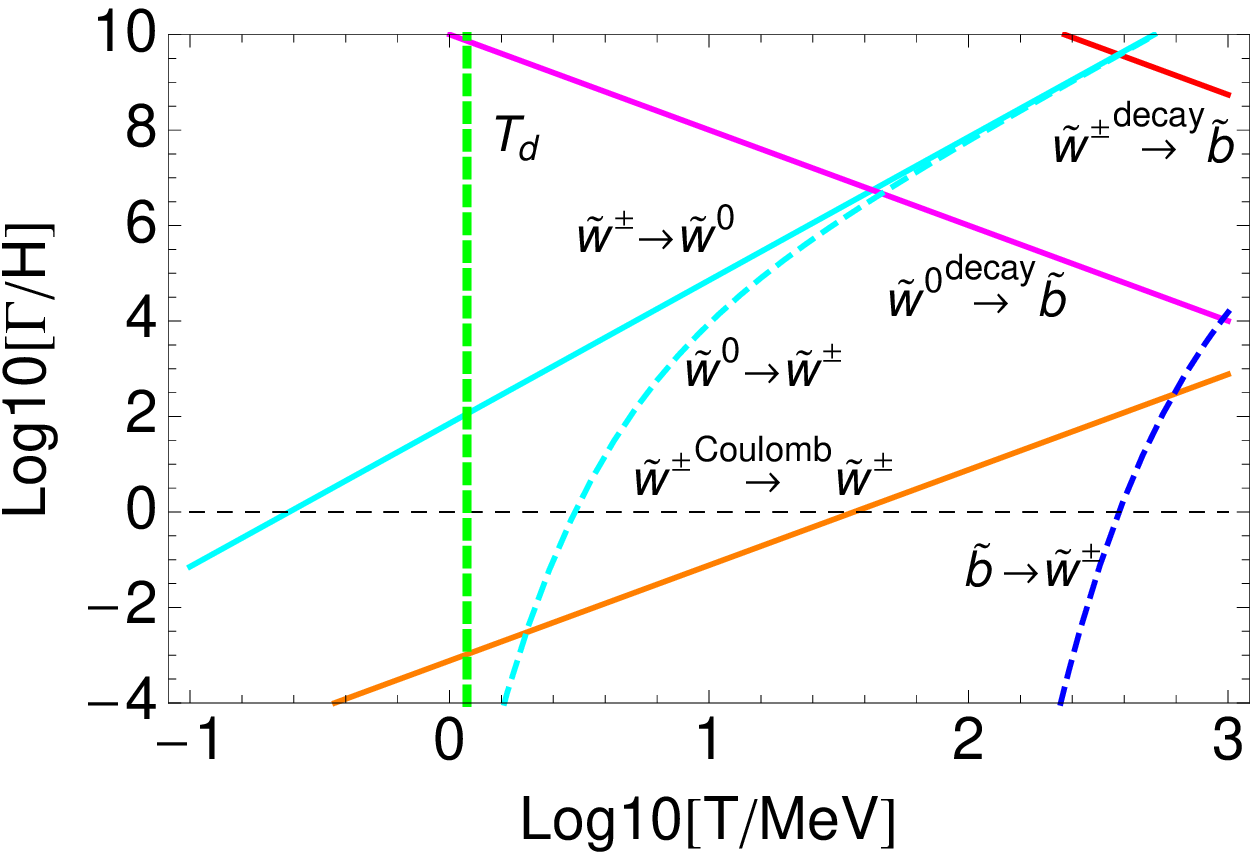}
  \end{center}
  \end{minipage}
 \begin{minipage}{.49\linewidth}
 \begin{center}
 \includegraphics[width=\linewidth]{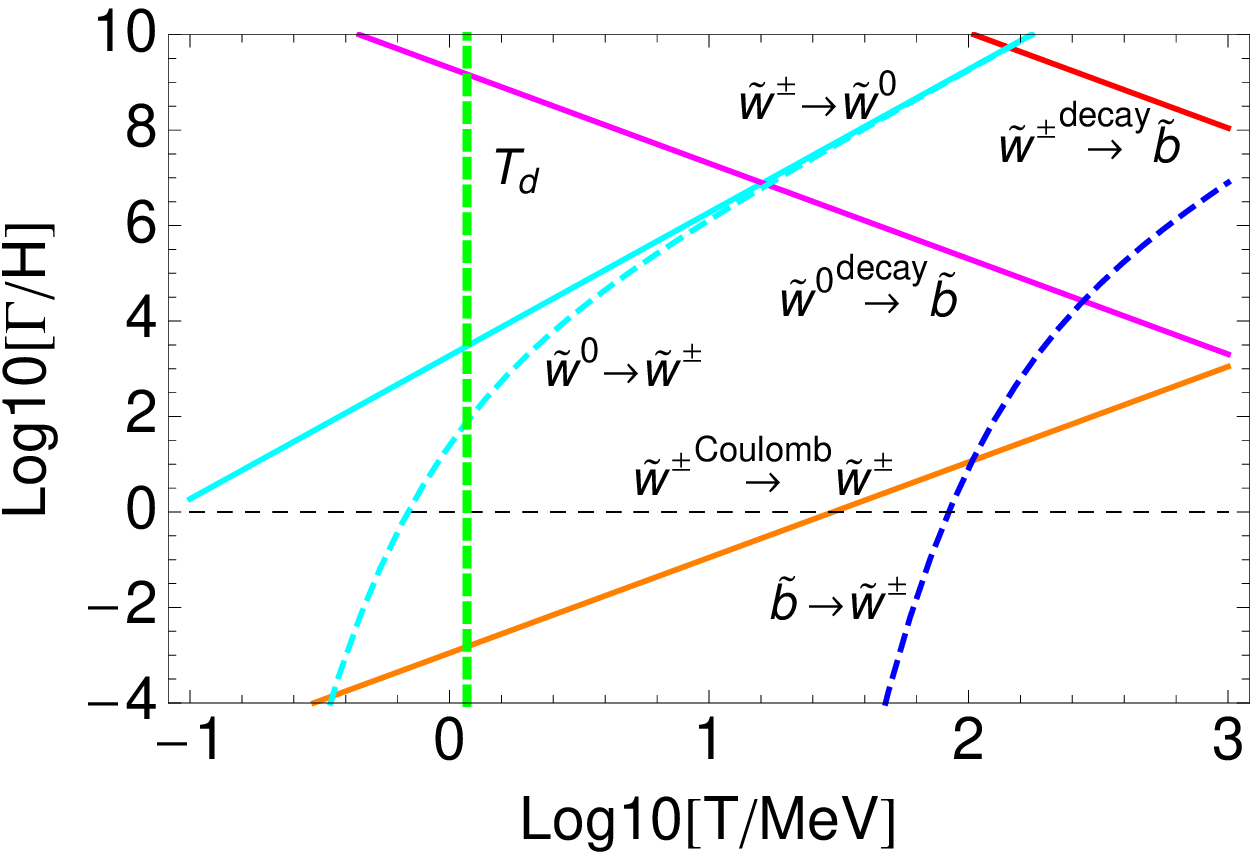}
  \end{center}
 \end{minipage}
\caption{\sl \small
The normalized reaction rates of interactions relevant to the energy-loss of the non-thermal bino (wino). Here, we take $m_{\tilde b} = 600$\,GeV and ${\mit\Delta} m_{\tilde w}=29.7$\,GeV ($\leftrightarrow$ $r_{\rm T}=0.5$) for $E_{{\tilde b}({\tilde w})} = 2$\,TeV (left panel) and $E_{{\tilde b}({\tilde w})} = 10$\,TeV (right panel). 
The vertical line shows the decay temperature of the gravitino (Eq.\,(\ref{eq:decaytemp})).}
\label{fig:kinetic12}
\end{figure}

The charged winos lose their energy efficiently via Coulomb scattering,
\begin{eqnarray}
-\frac{dE_{{\tilde w}^{\pm}}}{dt}=\frac{\pi \alpha^2 T^2}{3}\Lambda \left(1-\frac{m^2_{\tilde w}}{2E^2_{{\tilde w}^{\pm}}}\ln \left( \frac{E_{{\tilde w}^{\pm}}+p_{{\tilde w}^{\pm}}}{E_{{\tilde w}^{\pm}}-p_{{\tilde w}^{\pm}}} \right) \right) \,,
\end{eqnarray}
with the Coulomb logarithm $\Lambda$, which is estimated as 
\begin{eqnarray}
\Lambda = \ln \left[ \frac{4 \langle p_{e}^{2} \rangle}{k_{\rm D}^{2}} \right] \,, \quad
\langle p_{e}^{2} \rangle \sim \left( \frac{E_{{\tilde w}^{\pm}}}{m_{{\tilde w}^{\pm}}} T \right)^{2} \,, \quad
k_{\rm D}^{2} \simeq \frac{4 \pi \alpha}{3} T^{2} \,,
\end{eqnarray}
taking into account the contributions to the Debye screening scale $k_{\rm D}$ from the relativistic electrons and the relativistic positrons.
At the temperature of interest, $T_{\rm d} \simeq 0.1-10$\,MeV, the charged wino turns into the neutral wino mainly via the 2-body decay process (see Fig.\,\ref{fig:kinetic12}) in the case of the wino LSP.
In this case, the relatively long lifetime of the charged winos allow the non-thermal charged 
winos lose most of their energy before they decay\,\cite{winoenergylossibe}, 
\begin{eqnarray}
\tau_{{\tilde w}^{\pm}} \equiv \frac{1}{m_{{\tilde w}^{\pm}} \Gamma_{{\tilde w}^{\pm} \to {\tilde b} + f_{1} + {\bar f}_{2}}} \left( - \frac{dE_{{\tilde w}^{\pm}}}{dt} \right) \gg 1.
\end{eqnarray}
However, in the present scenario, the charged winos mainly decay into the bino LSP with a shorter lifetime.
For the mass splitting of ${\mit \Delta} m_{\tilde b} \gtrsim 5$\,GeV, the energy-loss of the charged winos becomes inefficient, $-dE_{{\tilde w}^{\pm}} / \left( E_{{\tilde w}^{\pm}} \Gamma_{{\tilde w}^{\pm}} dt \right) \lesssim 1$ , at $T \gtrsim 1$\,MeV.

The neural wino itself does not have elastic energy-loss processes at the tree-level.
The neutral wino lose its energy through the inelastic scattering of ${\tilde w}^{0} + e\,(\nu_{e}) \to {\tilde w}^{\pm} + \nu_{e}\,(e)$.
The inelastic scattering rate is given by,
\begin{eqnarray}
\label{eq:kinetic}
\Gamma_{{\tilde w}^{0}, \, {\rm inelastic}} &=& \frac{8}{\pi^3} G_{F}^2 T^5 \frac{(E_{{\tilde w}^{0}}+p_{{\tilde w}^{0}})^4}{m_{{\tilde w}}^2E_{{\tilde w}^{0}}p_{{\tilde w}^{0}}} \left(6+2\frac{m_{\tilde w}}{E_{{\tilde w}^{0}}+p_{{\tilde w}^{0}}}\frac{{\mit \Delta} m_{\tilde w}}{T} \right) \notag \\
&& \times \exp \left(-\frac{m_{\tilde w}}{E_{{\tilde w}^{0}}+p_{{\tilde w}^{0}}}\frac{{\mit \Delta} m_{\tilde w}}{T} \right)\,.
\end{eqnarray}
It should noted that this formula is applicable to only the relativistic wino and it is different from Eq.\,(\ref{eq:winononrelainelastic}) that is for the non-relativistic wino.
This formula can be easily translated into the inelastic scattering rate for the relativistic bino 
by multiplying the mixing and substituting physical quantities (e.g. mass splitting) related to the bino instead of the wino,
\begin{eqnarray}
\Gamma_{{\tilde b}, \, {\rm inelastic}} = N_{1 2}^{2} \Gamma_{{\tilde w}^{0}, \, {\rm inelastic}} ({\tilde w}^{0} \to {\tilde b})\,.
\end{eqnarray}
In Fig.\,\ref{fig:kinetic12}, we show the reaction rates for $m_{\tilde b} = 600$\,GeV and ${\mit\Delta} m_{\tilde w}=29.7$\,GeV, which corresponds to the case of $r_{\rm T}=0.5$.
The energy of the bino (wino) is different in each panel, $E_{{\tilde b} \, ({\tilde w})} = 2$\,TeV for the left panel and 
$E_{{\tilde b} \, ({\tilde w})} = 10$\,TeV for the right panel.
The inelastic scattering rate both for the bino and for the neutral wino sharply drops around
\begin{eqnarray}
\label{eq:criticaltemp}
T_{\rm c} = \frac{m_{{\tilde b} \, ({\tilde w})}  {\mit \Delta} m_{{\tilde b} \, ({\tilde w})}}{2 E_{{\tilde b} \, ({\tilde w})}} \simeq 900 \, {\rm MeV} 
\left( \frac{m_{{\tilde b} \, ({\tilde w})}}{600\,{\rm GeV}} \right) \left( \frac{{\mit \Delta} m_{{\tilde b} \, ({\tilde w})}}{30\,{\rm GeV}} \right)
\left( \frac{E_{{\tilde b} \, ({\tilde w})}}{10\,{\rm TeV}} \right)^{-1}\,,
\end{eqnarray}
due to the Boltzmann factor in Eq.\,(\ref{eq:kinetic}).

\begin{table}[tb]
 \caption{\sl \small
 The energy-loss cycle of the bino and the winos around and after the gravitino decay.}
 \label{table:cycle}
  \begin{center}
  \begin{tabular}{|c||c|} \hline
  particle & dominant process  \\ \hline \hline
  ${\tilde w}^{\pm}$ & ${\tilde w}^{\pm} \stackrel{\rm Coulomb}{\to} {\tilde w}^{\pm} \to {\tilde b} + f_{1} + {\bar f}_{2}$ \\ \hline
  ${\tilde w}^{0}$ & ${\tilde w}^{0} + e\,(\nu_{e}) \to {\tilde w}^{\pm} + \nu_{e}\,(e)$ (high T) \\ \cline{2-2}
  & ${\tilde w}^{0} \to {\tilde b} + f + {\bar f}$ (low T) \\ \hline
  $\tilde b$ & ${\tilde b} + e\,(\nu_{e}) \to {\tilde w}^{\pm} + \nu_{e}\,(e)$ \\ \hline
  \end{tabular}
 \end{center}
\end{table}

From Fig.\,\ref{fig:kinetic12}, we can identify the energy-loss cycle of 
the bino and the winos around and after the gravitino decay and summarize it in Table\,\ref{table:cycle}. 
The non-thermal charged winos ${\tilde w}^{\pm}$ lose their energy thorough the Coulomb interaction and then decay into the bino $\tilde b$.
The energetic bino $\tilde b$ is scattered inelastically and turns into the charged wino ${\tilde w}^{\pm}$. 
The relativistic neutral wino goes in two ways depending on the cosmic temperature. 
If the temperature is high enough, the inelastic scattering rapidly turns the neutral wino ${\tilde w}^{0}$ into the charged wino ${\tilde w}^{\pm}$.
Otherwise, it decays into the bino $\tilde b$ before it is scattered inelastically.

\subsection{Boltzmann equation and characteristic quantities}
The discussion in the previous subsection clarifies the evolution equation of the bino momentum spectrum that should be solved.
However, the calculation cost is still high and thus, we further simplify the evolution equation as follows without missing the essence.
First, we take into account the incomplete energy-loss of the charged winos until their decay by changing the bino inelastic scattering rate as,
\begin{eqnarray}
&& \Gamma_{{\tilde b}, \, {\rm inelastic}} \to \left( 1 - e^{- \tau_{{\tilde w}^{\pm}}} \right) \Gamma_{{\tilde b}, \, {\rm inelastic}} \,.
\end{eqnarray}
The prefactor $\left( 1 - e^{- \tau_{{\tilde w}^{\pm}}} \right)$ represents the probability of complete energy-loss at each inelastic scattering.
The second simplification is for the neutral wino process.
As we mentioned, the dominant process for the neutral wino shifts from the inelastic scattering to the decay as the temperature of the Universe decreases. 
We assume that this takes place instantaneously at the time
\begin{eqnarray}
\tau_{{\tilde w}^{0}} \equiv \frac{\Gamma_{{\tilde w}^{0}, \, {\rm inelastic}}}{(m_{{\tilde w}^{0}} / E_{{\tilde w}^{0}}) \Gamma_{{\tilde w}^{0} \to {\tilde b} + f + {\bar f}}} = 1 \,. 
\end{eqnarray}

With these simplifications, the Boltzmann equation of the momentum spectrum of the ``warm" bino dark matter, $f_{\rm warm}(p, t)$, can be written as,
\begin{eqnarray}
\label{eq:Boltzmanneq}
&& \frac{\partial}{\partial t} f_{\rm warm}(p, t) - H p \frac{\partial}{\partial p} f_{\rm warm}(p, t) \notag \\
&& \quad
= {\bigg [} e^{- \tau_{{\tilde w}^{\pm}}}
\frac{d\Gamma_{3/2, \, {\tilde w}^{\pm}}}{d^3 p} 
+ e^{ - \Theta (\tau_{{\tilde w}^{0}} - 1) \tau_{{\tilde w}^{\pm}}} \frac{d\Gamma_{3/2, \, {\tilde w}^{0}}}{d^3 p}
+ \frac{d\Gamma_{3/2, \, {\tilde b}}}{d^3 p} {\bigg ]}
\frac{a(t_{0})^3}{a(t)^3}e^{-\Gamma_{3/2} t} \notag \\
&& \qquad
- \left( 1 - e^{- \tau_{{\tilde w}^{\pm}}} \right) \Gamma_{{\tilde b}, \, {\rm inelastic}} \, f_{\rm warm}(p, t) \,,
\end{eqnarray}
with the Heaviside step function $\Theta(x)$.
The differential decay rate of the gravitino into the bino $\tilde b$ (wino ${\tilde w}$), 
$d\Gamma_{3/2, \, {\tilde b} \, ({\tilde w})} / d^3 p$, are given in Ref.\,\cite{winoenergylossibe}.
The prefactor $e^{ - \Theta (\tau_{\tilde b} - 1) \tau_{{\tilde w}^{\pm}}}$ and $e^{- \tau_{{\tilde w}^{\pm}}}$ represent
the energy-loss of the neutral and the charged winos immediately after their production, respectively.
The momentum spectrum of the ``warm'' bino dark matter is normalized such that,
\begin{eqnarray}
\int \frac{d^3 p}{(2 \pi)^{3}} \, f_{\rm warm}(p, t) {\Big |}^{\tau_{{\tilde w}^{\pm}}=0}_{t=t_{0}} = 1 \,,
\end{eqnarray}
at present $t=t_{0}$ when we turn off the energy-loss process by hand, $\tau_{{\tilde w}^{\pm}}=0$.

After obtaining the momentum spectrum of the ``warm" bino dark matter, we calculate two quantities, 
which characterize the ``warmness'' of dark matter (see Ref\,\cite{winoenergylossibe} for the case of the wino LSP).
One is the resultant ``warm'' fraction of dark matter,
\begin{eqnarray}
r_{\rm warm} = (1-r_{\rm T}) \int \frac{d^3 p}{(2 \pi)^{3}} \, f_{\rm warm}(p, t) {\Big |}_{t=t_{0}} \,.
\end{eqnarray}
The larger ``warm'' fraction leads to the more suppression of the matter power spectrum.
The other is the free-streaming scale, which is defined by the Jeans scale at the matter radiation equality $a_{\rm eq}$,
\begin{eqnarray}
&&k_{\rm fs}= a \sqrt{\frac{4 \pi G \rho_{\rm mat}}{\langle v^2 \rangle}} {\bigg |}_{t=t_{\rm eq}} \,, \notag \\
&&\langle v^2 \rangle (t_{\rm eq}) = (1-r_{\rm T}) \frac{a(t_{0})^{2}}{a(t_{\rm eq})^{2}} \int \frac{d^3 p}{(2 \pi)^{3}} \, \frac{p^{2}}{m_{\tilde b}^{2}} f_{\rm warm}(p, t) {\Big |}_{t=t_{0}} \,.
\end{eqnarray}
The free-streaming scale determines the critical scale below which the suppression on the matter power spectrum becomes significant\,\cite{warmkamada}.

\subsection{Results}

\begin{figure}[tb]
 \begin{minipage}{.49\linewidth}
 \begin{center}
 \includegraphics[width=\linewidth]{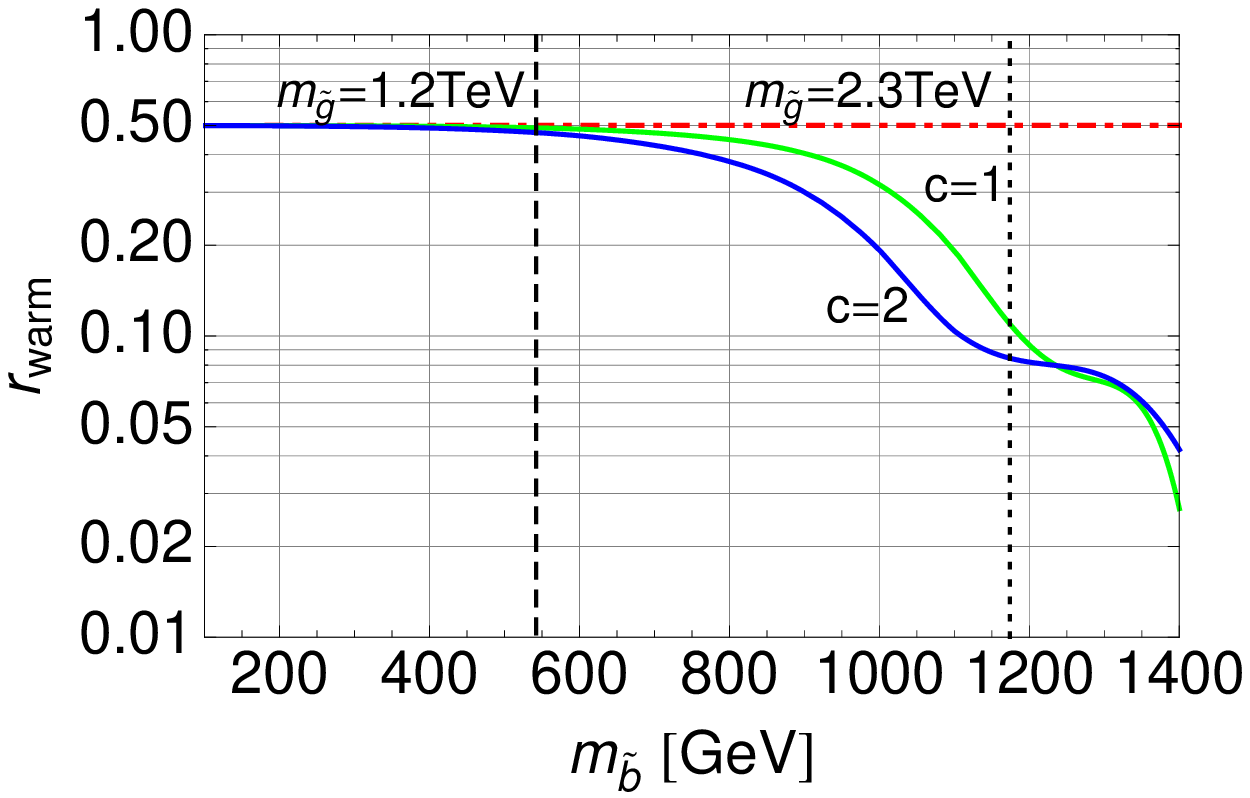}
  \end{center}
  \end{minipage}
 \begin{minipage}{.49\linewidth}
 \begin{center}
 \includegraphics[width=\linewidth]{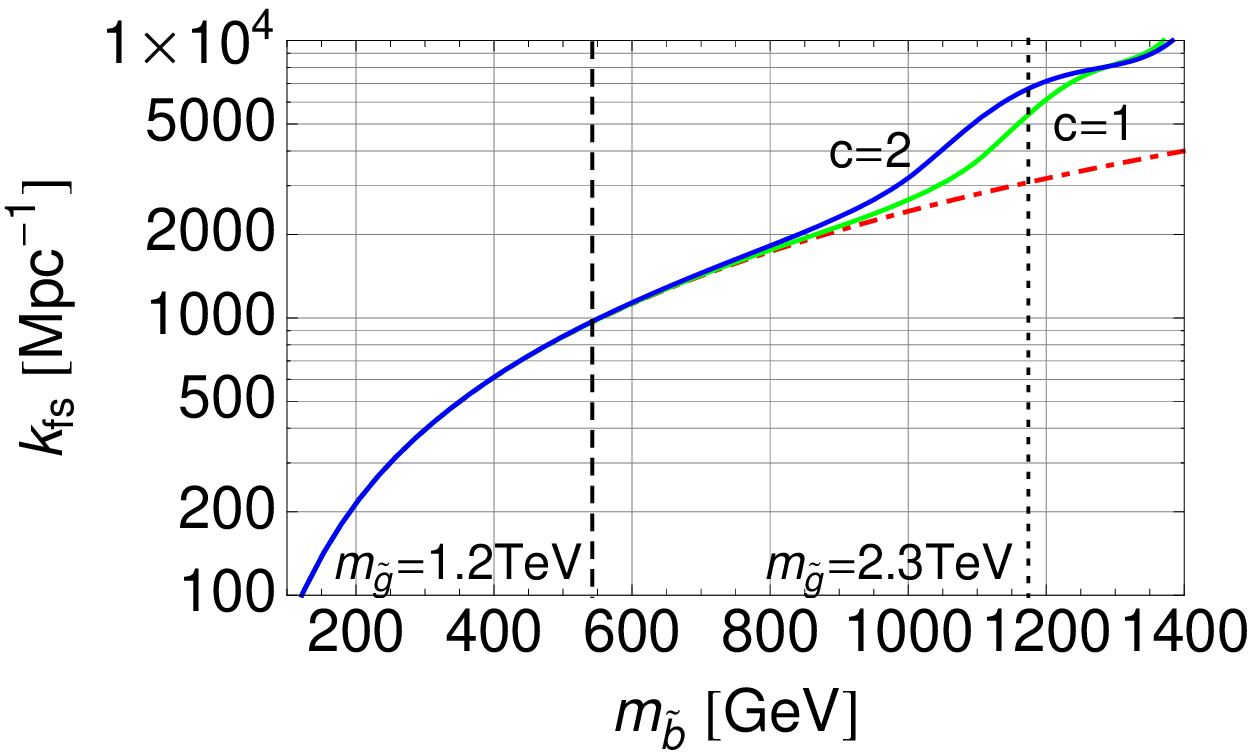}
  \end{center}
 \end{minipage}
\caption{\sl \small
The ``warm'' fraction $r_{\rm warm}$ (left panel) and the free-streaming scale $k_{\rm fs}$ (right panel).
We set the mass splitting such that $r_{\rm T}=0.5$ (see Fig.\,\ref{fig:rt}).
The different choice of $c\,(=1,\,2)$ corresponds to the different value of the sfermion mass and the higgisino mass, $m_{\tilde f}=\mu=c\,m_{3/2}$.
For comparison, we plot the ``warmest'' case ($\leftrightarrow \tau_{{\tilde w}^{\pm}}=0$) in the dot-dashed line.
Here, we also show the constraint on the bino mass from the current (8\,TeV) and the future (14\,TeV) gluino search at the LHC.}
\label{fig:warm05}
\end{figure}

\begin{figure}[tb]
 \begin{minipage}{.49\linewidth}
 \begin{center}
 \includegraphics[width=\linewidth]{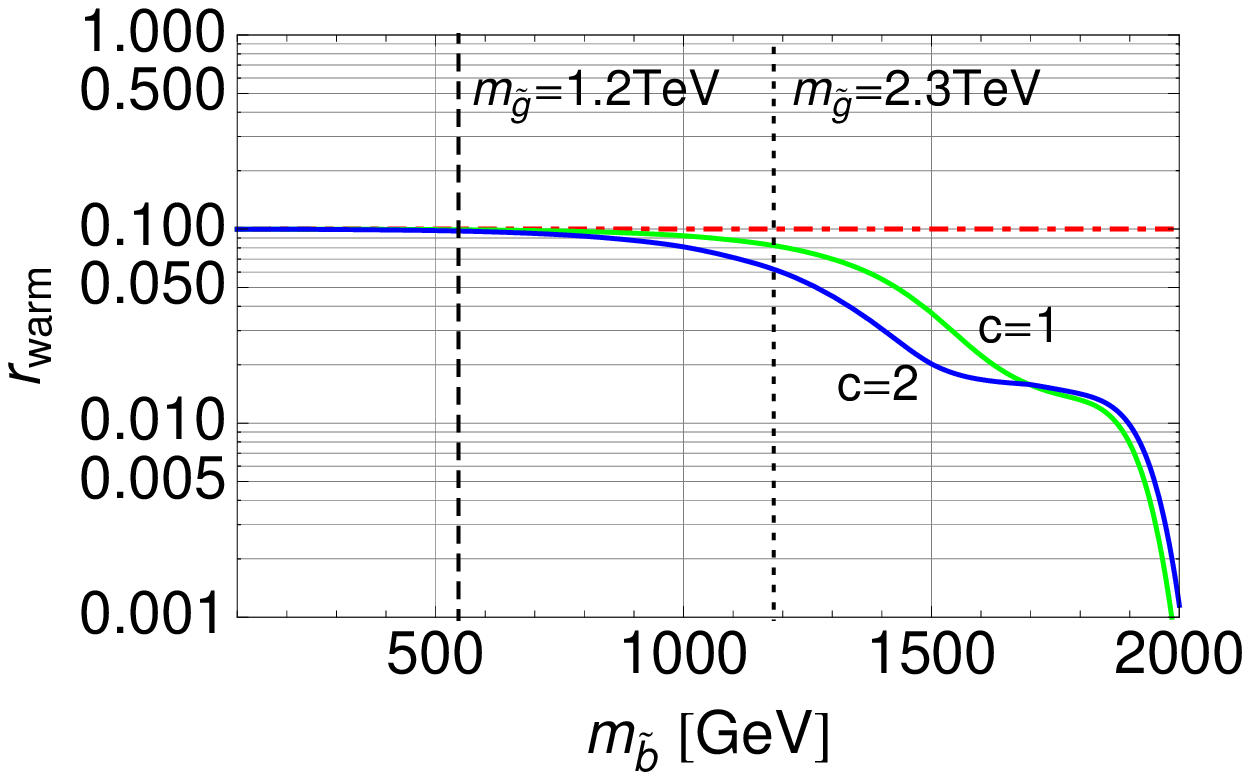}
  \end{center}
  \end{minipage}
 \begin{minipage}{.49\linewidth}
 \begin{center}
 \includegraphics[width=\linewidth]{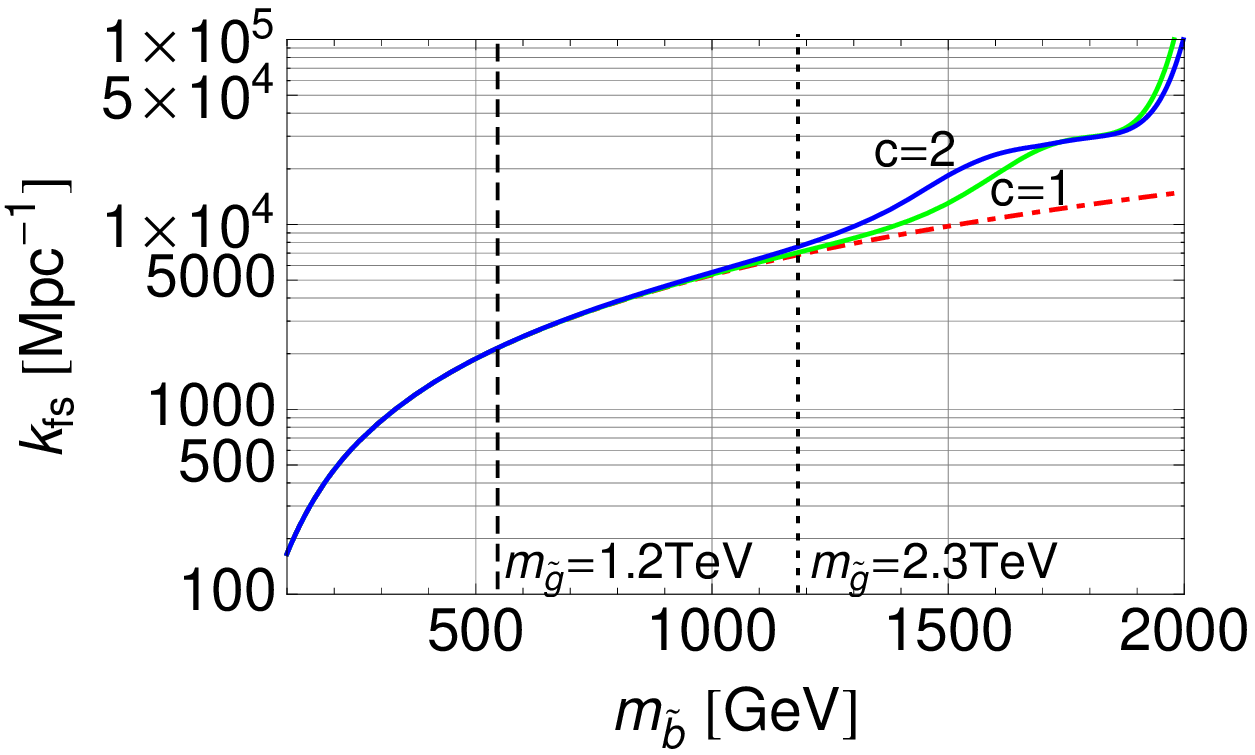}
  \end{center}
 \end{minipage}
\caption{\sl \small
The same plot as Fig.\,\ref{fig:warm05}, but for $r_{\rm T}=0.9$.}
\label{fig:warm09}
\end{figure}

For each bino mass $m_{\rm b}$, we solve the Boltzmann equation (Eq.\,(\ref{eq:Boltzmanneq})) numerically to calculate $r_{\rm warm}$ and $k_{\rm fs}$.
The results are shown in Fig.\,\ref{fig:warm05} ($\leftrightarrow r_{\rm T} = 0.5$) and Fig.\,\ref{fig:warm09} ($\leftrightarrow r_{\rm T} = 0.9$).
The suppression of $r_{\rm warm}$ for the heavier bino mass $m_{\tilde b}$ is owing to the larger gravitino mass.
The larger gravitino mass has two effects.
First, the heavier gravitino decays in the earlier and hotter Universe, where the energy-loss processes are more efficient.
Second, the bino and the winos are more energetic at the decay of the heavier gravitino, 
for which the inelastic scatterings are less suppressed.

While we set $m_{\tilde f}=\mu=m_{3/2}$ in the above discussion, this relation can be different by an order of one factor.
In order to take into account this ambiguity, we introduce the parameter $c$, which is defined by
\begin{eqnarray}
m_{\tilde f}=\mu=c\,m_{3/2} \,.
\end{eqnarray}
The effects of $c$ parameter can be interpreted as follows.
For the heavier sfermions and the heavier higgsinos ($c=2$), the inelastic scattering rate $\Gamma_{{\tilde b}, \, {\rm inelastic}}$ is more suppressed.
However, as can be seen from Fig.\,\ref{fig:kinetic12}, the inelastic scattering rate has dropped around the critical temperature (Eq.\,\ref{eq:criticaltemp}) 
well above the decay temperature of the gravitino $T_{\rm d}$ (Eq.\,\ref{eq:decaytemp}).
Therefore the additional order of one factor $c$ does not have significant effects through the inelastic scattering rate.
On the other hand, the energy-loss via Coulomb scattering remains efficient until just before the gravitino decay.
The large $c$ prolongs the lifetime of the charged winos and thus, enhances their energy-loss during one lifetime $\tau_{{\tilde w}^{\pm}}$.
Moreover, the large $c$ increases the fraction of the non-thermal neutral wino that turns into the charged wino (i.e. large $\tau_{{\tilde w}^{0}}$).
Therefore, the large $c$ results in the ``colder'' bino dark matter with the smaller $r_{\rm warm}$ and the larger $k_{\rm fs}$.

Before closing this subsection, we remark the implication of our results.
As we can see in Fig.\,\ref{fig:warm05} and Fig.\,\ref{fig:warm09}, in the favorable (not constrained) parameter region,
the ``warm'' component accounts for sizable fraction (at least $1$\,\%) of the whole bino dark matter.
This is why we refer the bino dark matter to mixed (cold+warm) dark matter in the present model.
The bino dark matter can not resolve the so-called ``small scale crisis'', 
since the free-streaming scale should be much smaller for that purpose, e.g. $k_{\rm fs} \simeq 20-200 \, {\rm Mpc}^{-1}$\,\cite{warmkamada}.
Its imprints on the small-scale matter power spectrum, however, are significant before the formation of non-linear objects (e.g. dark matter halos) in the Universe.
The future $21$\,cm survey will probe matter density fluctuations at high-redshifts, e.g. $z \simeq 30-200$, and thus provide us an important hint
on the non-thermal production of the bino dark matter.


\section{Summary}
\label{sec:summary}
In this paper, we studied the bino-wino co-annihilation scenario in high-scale SUSY breaking models with the heavy sfermions and the heavy higgsinos.
As one specific realization, we consider the pure gravity mediation/minimal split SUSY model, which is highly motivated after the discovery of the Higgs boson.
The wino LSP is now in tension with indirect dark matter searches by the Fermi-LAT and the H.E.S.S. telescope,
while there is still large ambiguity in the dark matter profile at the Galactic center. 
On the other hand, the bino LSP is almost free from any direct and indirect detections.

The suppressed interaction of the bino dark matter generically results in the over-closure of the Universe.
In order to account for the observed dark matter density, the bino LSP should be accompanied by the slightly heavier wino NLSP.
The small mass splitting between the bino and the wino allows the sizable amount of the winos to exist at the freeze-out of the bino and 
it boosts the annihilation effectively. 
Therefore, for the first step, we identify the mass splitting needed for a correct bino thermal relic. 
In the calculation of the thermal relic, we take into account both the co-annihilation and the Sommerfeld enhancement.
For that purpose, we also clarify the relevant processes in charge of keeping chemical equilibrium between the bino and the winos.

For smaller mass splittings, the bino thermal relic can not account for the whole dark matter density.
We assume the late time decay of the gravitino produces the non-thermal bino after the freeze-out.
The non-thermal bino is produced with sufficiently high energy to be ``warm'' dark matter.
However, the ``warmness'' of the bino dark matter depends on the energy-loss processes after the production.
To this end, we clarified the energy-loss cycle of the non-thermal bino and the non-thermal winos.
With several reasonable simplifications, we derive the Boltzmann equation of the momentum spectrum of the ``warm'' bino dark matter.
We solve it numerically and show that the ``warm'' component accounts for sizable fraction (at least $1$\,\%) of the bino dark matter.
The matter power spectrum is suppressed below $1-10$\,kpc ($\leftrightarrow k_{\rm fs}=10^{3}-10^{4} \, [{\rm Mpc}^{-1}]$). 
As a result, we find that the imprints of non-thermal component on the small-scale matter power spectrum provides an invaluable hint
on the present scenario, e.g.  in the future $21$\,cm surveys. 

In this paper, we concentrated on the pure gravity mediation/minimal split SUSY model and its gaugino mass spectrum (Eq.\,(\ref{eq:gluinomass})-(\ref{eq:binomass})).
However, the existence of the extra vector-like matter fields can change the gaugino mass spectrum\,\cite{extendedpuregrav}.
In this case, the gluino mass as well as the wino mass can be close to the bino mass.
The mass degeneracy hides the gluino from the collider experiments, since the decay products of the gluino do not have sufficient energy to be distinguished from the background events.
On the other hand, the cosmological imprints, which are discussed in this paper, can be enhanced.
This is because the gravitino mass can be smaller for the fixed bino mass.
The lighter gravitino leaves the non-thermal bino with large comoving velocity at the gravitino decay (Eq.\,(\ref{eq:comovingvb})).
Moreover, the lighter gravitino decays at very late time (still before BBN), at which the energy-loss processes are insufficient (see Fig.\,\ref{fig:kinetic12}).

\section*{Acknowledgments}
This work is supported by Grant-in-Aid for Scientific research from the Ministry of Education, Science, Sports, and Culture (MEXT), Japan, No. 24740151 (M.I.), No. 22244021 and No. 23740169 (S.M.), JSPS Research Fellowships for Young Scientists (A.K.) and also by World Premier International Research Center Initiative (WPI Initiative), MEXT, Japan.


\end{document}